# A novel approach to quantify the structural distortions of U/Th snub-disphenoids and their role in zircon → reidite type phase transitions of uranothorite

Sudip Kumar Mondal[1,2,a] Pratik Kr. Das[3,b], Nibir Mandal[2,c], A. Arya[4,d]
[1] Department of Physics, Jadavpur University, Kolkata 700032, India.
[2] High Pressure and Temperature Laboratory, Faculty of Science, Jadavpur University, Kolkata 700032, India.
[3] The Centre for Earth Evolution and Dynamics, University of Oslo, Oslo, N-0315, Norway.
[4] Material Science Division, Bhabha Atomic Research Centre, Mumbai, 400085, India.

Using ab initio calculations in the framework of density functional theory (DFT) we have investigated the structural stability and mechanical properties of Uranothorite ($U_{1-x}Th_xSiO_4$) (x= 1.0, 0.75, 0.5, 0.25, 0.0). These actinide orthosillicate phases have a kaleidoscopic range of applications, starting from eminently functional materials in nuclear industries, such as fission reactor development, storage and disposal of spent nuclear fuels, and vitrified nuclear wastes, to geochronological dating materials. The questions we pose here pertains three crucial issues- 1) how does incorporation of Th in place of U modify the crystal structure and mechanical behavior; 2) how can such elemental substitution affect the structural transition from zircon (Z)-type ($I4_1/amd$) to reidite (R)-type ($I4_1/a$) phase; and 3) what is the influence of the coordination geometry of the actinide polyhedra on the transition pressure?

Our DFT calculations show the incorporation of Th in place of U enforces a linear increment in unit-cell volume in both phases. But, the traits of facilitating the volume increment in the Z and R phases markedly differ from each other. This analysis yields a peculiar relation of the Z→R transition pressure ($p_t$) with normalized Th content i.e. Th/(U+Th). For coffinite, $USiO_4$, $p_t$ is found to be 8.54 GPa, which drops to 7.67 GPa when Th/(U+Th) = 0.25 and attains a minimum of 6.82 GPa with equal amount of U and Th, i.e., Th/(U+Th) = 0.50. Further increase in Th concentration reversed the trend of $p_t$ variation, and gave rise to $p_t$ = 7.92 GPa and 8.68 GPa for Th/(U+Th) = 0.75 and 1.0, respectively. The R-type phases continue to enhance their compressibility with increasing Th content, whereas the Z-type phases attains a maximum value of 5.97 x $10^{-3}$ $GPa^{-1}$ for Th/(U+Th) = 0.50, and then starts to decline with increasing Th content.

Earlier studies have estimated the distortion in coordination polyhedra, but their calculations were limited to platonic solids. But, the uranothorite phases display strongly irregular polyhedral structures, where the 8 coordinated U/Th atoms constitute triangular dodecahedron or snub-disphenoid with 8 O atoms at their vertices. Out of eight atoms, four of them, however, form the edge by assembling with 4 nearest neighbor O atoms, and the rest 4 assemble with 5 nearest neighbor O atoms to form edges of different geometry. These contrasting combinations lead to two types of U/Th-O bonds of strikingly different lengths and angles with multiplicity 2, 4 and 12. We theoretically define a parameter ($L$) as the ratio of two configurationally different bond lengths of undistorted polyhedra. The value of $L$ is found to unique (~1.374302), independent to the polyhedral size of the snub-disphenoid. Departure ($\delta$) from this unique value is accounted as a measure of bond length distortion. Similarly, we have used the sample variance of bond angles to quantify their distortions. The bond angle variances of U and Th-polyhedra show a maximum value for Z-phase, whereas a minimum limit for the R-phase at Th/(U+Th) = 0.50. For this concentration, $\delta$ attains a maximum in case of the R-phase. To conclude, the Z - R phase transition pressure, $p_t$ declines to a minimum when U and Th occur equally in the uranothorite phase, maintaining the least differences in polyhedral distortions.

[a] sudipkm1990@gmail.com
[b] pratik@geo.uio.ni
[c] nibirmandal@yahoo.co.in
[d] ashokarya@gmail.com

## I. Introduction

Coffinite ($USiO_4$), its Thorium containing counterpart Uranothorite ($U_{1-x}Th_xSiO_4$) and Thorite ($ThSiO_4$) are conspicuous members of the nesosilicate subclass of actinide silicates. They are isostructural with Zircon and Hafnon. Coffinite, accompanied by Uranothorite solid solutions are major phases frequently encountered as microcrystals in uranium ores[1] which are economically suitable for radiogenic uranium extraction, with the former appearing to be the second most abundant U(IV) mineral on earth. But they exhibit significant differences in physical and chemical environment of formation. Coffinite prefers to precipitate in low temperature hydrothermal/sedimentary condition, whereas the later prefers a high temperature hydrothermal magmatic condition[2]. Although Thorite and its high pressure-temperature monazite structured polymorph Huttonite occur naturally, they are a bit scarce relative to Coffinite. However both are found as accessory minerals in medium grained metamorphic rocks like schists[3] and sometimes in unusually coarse grained plutonic igneous rocks i.e. in pegmatite, crystallized from magma during the ultimate stage[4,5]. These orthosilicates along with Zircon and Hafnon have a kaleidoscopic range of utilizations. Due to their highly effective corrosion resistivity, chemical stability and radiation resistance they are suitable for straight disposal of spent nuclear fuel (SNF) and vitrified nuclear waste[6,7] in underground repositories, which is evidently an emerging and daunting scientific challenge of modern era. For the same reasons, they are ideal for applications in fission reactors. Although they suffer structural damages due to autoradiation, the reliability and accuracy achieved in recent experimental aspects, makes them legitimate candidate for radiogenic geochronology[2,8,9]. For possessing a high resistance to deformation and a wide range of stability over upper mantle conditions the zircon structure is considered to be a reliable host phase for minerals of deep mantle origin. $U^{238}$ and $Th^{232}$ are prolific sources of geoneutrinos. The results of KamLAND collaboration estimates that radioactive decay of $U^{238}$ and $Th^{232}$ contribute to 20TW to earths heat flux[10]. As U and Th are the heaviest lithofile elements, their silicate phases are also executive in producing radiogenic heat in shallow interiors of the earth. Consequently, an account of distribution of their sources and their physical behavior is a route to explore the thermal evolution of the dynamic earth.

There is a significant amount experimental study focusing on the synthesis of radioactive orthosilicates and the thermochemical problems encountered (e.g. Crystallization temperature, pH of solutions used, solubility of U/Th in aqueous solutions, heating time, U/Si molar ratio calorimetric issues etc.) there in[11–16]. Guo et al. have experimentally demonstrated that the energetic metastability of coffinite with respect to quartz and uraninite ($UO_2$) mixture makes it impossible to synthesize the former from the later mixture. They concluded that the alternative way is to go for the precipitation in hydrothermal environment in low temperature. Therefore, provided that the formation of coffinite can proceed through precipitation from aqueous solution, it can be an alteration product of $UO_2$ subjected to hydrothermal, metamorphic, or even igneous conditions[13]. However, the previous studies mainly concern temperature as the significant thermodynamic variable. But in order to act as an underground storage of SNF the orthosilicates have to circumvent chemical decomposition in response to larger pressures too. Zircon is subjected to extreme conditions due to tectonic activities in lithospheric mantle; one such environment is where kimberlite eruption takes place, and additionally during shock metamorphism owing to hypervelocity meteoritic impact in our planet or other planetary bodies[17]. This mineral phase is a competent one to provide archive of different extreme pressure (p)-temperature (T) regimes that it had experienced and thus, considered to be a prospective phase for high p-T thermobarometry. The mineral zircon, its equation of state and high pressure phase transition to reidite/scheelite has been investigated experimentally by Reid and Ringwood[18] and Ono et al.[19]. A number of known Arsenate and Vandate compounds favor to undergo a pressure-induced phase transition from Zircon (Space Gr.- $I4_1/amd$) to reidite (Space Gr.- $I4_1/a$) type structures[20–24]. Similar phase transition under hydrostatic and dynamic stresses at ambient and high temperature for $ZrSiO_4$ and the physical properties of zircon and its high pressure polymorph have been rigorously investigated both experimentally and computationally[17,25–31]. Smirnov et al.[31], in his 'bond switching' mechanism, proposed that this phase transition is reconstructive in nature. Accumulation of shear elastic strains during transition causes the part of Zr-O bonds and the $ZrO_8$ polyhedra to disrupt and new Zr-O bonds form. The transition manifest itself via an intermediate transient cubic phase. Zhang et al[32] recently reported an identical transition for $USiO_4$ to occur at 14~17 GPa and proposed that the effect of pressure may also aid an electron transfer ($U^{4+}$ to $U^{5+}$) . Bauer et al.[33] revisited this transition and established it to be thermodynamically reversible. They suggested that softening of a silent vibrational Raman mode is responsible for it. The lattice dynamical study of Bose et al.[25] predicted that alike transition is possible for $ThSiO_4$ at pressure > 3 GPa. Nevertheless, there exists a paucity of complete description of the ambient physical properties and high pressure response of the members of continuous solid solution between $ThSiO_4$ and $USiO_4$. To the best of our knowledge, for the first time our result demonstrates a pressure induced anomalous first order zircon→reidite transition. This finding is further made evident by virtue of other physical alterations that we observed associated with it. It also asserts the existence of a solid solution between zircon type $USiO_4$ and $ThSiO_4$ which has been predicted by Goldschmit[34]. Our results also predict the possibility of a solid solution between $USiO_4$ and $ThSiO_4$ in reidite type phases.

Fluctuations in composition and/or thermodynamic parameters can introduce significant amount of geometric distortions in coordination polyhedra of unit cells. These distortions manifest themselves through convoluted behavior of physical properties. Numerical estimates of such contortions are therefore often solicited to explain the response of the crystals subjected to different alterations. Existing literature in this regard is comprised of two predominant methods. Robinson et al.[35] proposed that the degree of distortion of coordination polyhedron can be quantified by invoking two parameters, characteristic of the structures under study: quadratic elongation (λ) and angular variance$(\sigma_\theta^2)$. In an alternative way, Makovicky et al.[36] proposed a measure of distortions based on the ratio of volumes of the circumscribed spheres and of the polyhedra calculated, respectively for the real and ideal polyhedra of the same number of coordinated atoms having the same circumscribed spheres. However, both the ideas and quantification are limited to an extent in terms of their applicability. Robinson's consideration provides with very good numerical estimates for two cases of regular, convex and uniform polyhedron (tetrahedra (CN = 4), octahedra (CN = 6)) and can be generalized to cube, dodecahedron (CN =8) and icosahedron (CN = 12) i.e. to the rest three Platonic solids. Macovicky's way transcends a bit further and incorporates estimates of the distortion behavior of four polyhedrons with CN = 12 viz. cuboctahedron, anticuboctahedron, icosahedron and maximum volume hexagonal prism. Nevertheless, the instances where the polyhedra are either not regular or not uniform, or both are left unattended by these two existing methods. An interesting way of characterizing the distortions of $ZrO_8$-triangular dodecahedra is proposed by Mursic et al.[30]. He suggested that one can think of the $ZrO_8$-polyhedron as two interpenetrating $ZrO_4$-tetrahedra, distinguished by the manner of their geometrical attachment with the adjacent $SiO_4$ tetrahedra: edge sharing or corner sharing (one prolate and one oblate conformation). Later Marques et al.[28] treated the geometry of $ZrO_8$-snub disphenoid in Zircon and Reidite phases as prescribed by Mursic et al.[30] and obtained the measure of distortion following Robinson et al. But in case of Reidite phase the $ZrO_8$-polyhedra are linked with $SiO_4$-tetrahedra only through sharing corner i.e. O-atom and consequently the structural trait to distinguish between two $ZrO_4$-tetrahedra is absent in this high-pressure polymorph of Zircon (Fig. 1). So, the generalization of the idea of Mursic et al. to the cases of Reidite phases lacks geometrical rigor. In our study we propose an entirely new simple and elegant approach to quantify the longitudinal and angular distortion of 8-coordinated sunb-disphenoid by keeping its original geometry. We theoretically established the existence of a parameter as the square of the deviation from the ratio of lengths of two configurationally different bonds, the value of which is unique, regardless of the size of the snub-disphenoid. Any deviation from that unique value then finds the utilization as the numerical estimate of structural distortion.

The paper has been organized in the following way. Section-II presents the procedures of our first principles calculations. Section -III provides our results in three subsections. Section-III-A is dedicated to present the structural traits and parameters of the different members $U_{1-x}Th_xSiO_4$ at ambient condition and how the exchange between U-Th stoichiometry does effects them. This section is concluded with equation of state (EOS) description of zircon type $ThSiO_4$ and its high pressure reidite type polymorph. Section-III-B describes the effect of hydrostatic pressure on the anomalous structural transitions and gives computational estimates of the pressure of transitions and bulk moduli. The bulk moduli and compressibility of the zircon type phases were seen to show a peculiar alteration with varying U-Th atomic percentage. Section-III-C branches out to two mini subsections. The first one presents a brand new mathematical approach to quantify the distortions of triangular dodecahedra in general and consequently the theoretical establishment of two parameters. The next subsection is committed to elucidate the effect of these distortions of U-Th polyhedra on the phase transitions and mechanical properties. Finally, we conclude our findings in Section-IV.

# II. Computational Methodology

We have used Density functional theory to predict the variation of physical properties of Uranothorite solid solution as implemented in open-source Quantum ESPRESSO suite[37–39]. First principle calculations including spin-polarization were performed using projecter augmented-wave (PAW) method[40]. The exchange-correlation effects of the electrons were treated using GGA and the exchange correlation functional were chosen following Perdew et al.[41]. For U, Th, Si, O atoms the following orbitals were treated as valence states: U ($6s^2 7s^2 6p^6 6d^{1.5} 5f^{2.5}$), Th ($6s^2 7s^2 6p^6 6d^1 5f^1$), Si ($3s^2 3p^2$), O ($2s^2 2p^2$). The core cut-off radii for them are 2.1 a.u., 2.1 a.u., 1.9 a.u. and 1.1 a.u. respectively. The remaining core electrons along with the nuclei were treated by scalar relativistic PAW pseudopotentials incorporating a non-linear core correction, which were generated by Andrea dal Corso[42] using the 'Atomic' code following the Troullier-Martins pseudization scheme. The kinetic energy cut-off for each individual members of the solid solution was kept at 1700 eV. The BFGS (Broyden–Fletcher–Goldfarb–Shanno)[43,44] algorithm was used for geometrical optimization to find the ground state electronic structure under strict convergence criteria. In both the cases the convergence threshold for energy and forces were set to $4x10^{-10}$eV/atom and $4x10^{-8}$ eV/Å, respectively. For the Zircon and Reidite phases, the Brillouin zones were sampled by dense 7x7x8 and 7x7x4 Monkhorst-Pack[45] k-point grid respectively, which gave rise to 50 and 99 irreducible k-points in their brillouine zone.

However, in general LDA/GGA fails to describe the effects of strongly correlated *d* and *f* electrons of transition metals and *f*-block elements accurately. As a remedy, a Hubbard-U factor in terms of strong Coulomb-like Hartree-Fock electrostatic potential is introduced to the system to handle *f* and *d* electrons of U and Th. So, a set of simulations for the $U_{0.5}Th_{0.5}SiO_4$ configuration were performed with Hubbard-U varying from 1 to 5 eV, for both Uranium and Thorium. But no changes were observed in the orbital occupancy. It also underestimated the lattice constants which are incommensurate with experimental data. Thus, the Hubbard-U were not accounted for our further simulations i.e. the addition of an on-site Coulomb repulsion for the *d*-or *f*-electrons, have not been introduced.

## III. Results

### A. Crystal Structures in Ambient Conditions and EOS

The tetragonal unit cells of coffinite, thorite and compositionally varying members contain 24 atoms in their conventional cell, both in zircon and reidite phases. Both the phases have 4 formula units per conventional cell. The former phase exists in Laue group 4/mmm, exhibiting $I4_1/amd$ space group whereas the later exists in Laue group 4/m manifesting in space group $I4_1/a$. The positions of the Si and U/Th atoms are fixed by symmetry: they are located at (0, 1/4, 3/8) and (0, 3/4, 1/8) on the 4b and 4a Wyckoff sites, respectively. The 16h Wyckoff sites (0, u, v) are populated by the O-atoms, where u and v are internal parameters. We are restricting our discussion as extensive studies have already been done on the crystal structure of zircon and reidite[25,26,30,46]. In comparison with them the structures of coffinite, thorite and their high pressure phases are not numerous in literature. Interestingly, coffinite or thorite structure is characterized by a chain of alternating edge sharing $SiO_4$ tetrahedra and (U/Th)-$O_8$ triangular dodecahedra extending parallel to *c* crystallographic axis (Fig. 1). While in reidite phases the actinide polyhedra and $SiO_4$ tetrahedra are observed to be distributed in a zig-zag manner along *c*-axis when viewed perpendicular to *bc* plane (Fig. 1).

During transition from zircon to reidite type the mode of sharing between $SiO_4$ and U/$ThO_8$ polyhedra undergoes a significant modification. The reidite structure consists of two intercalated diamond lattice sites, one occupied with U and/or Th and the other with Si. They are coordinated with eight and four oxygen atoms, forming U/$ThO_8$-dodecahedra and $SiO_4$-tetrahedra, respectively. However, there are pronounced dissimilarities in terms of the connection between U/Th-dodecahedras and their nearest neighbor Si-tetrahedra. In zircon phase we can detect that both edges and corners are shared between the two types of cationic polyhedral to form a compact tetragonal cell. Whereas in their high pressure reidite type phases the edge sharing is observed between neighboring actinide polyhedra in contrast to corner sharing which is present between actinide polyhedral and $SiO_4$ tetrahedra (Fig. 1).

Our calculated values of lattice parameters, *c/a* ratios and volumes are presented in Table-I and compared with previous theoretical and experimental studies. In case of coffinite we found $a$ = 7.0303 Å and $c$ = 6.2872 Å which have a deviation of 0.65% and 0.5% from experimental[11] and theoretical study[33], respectively. They are also in excellent agreement with reported values of Zhang et al., Guo et al. and Pointeau et al. [12,15,32]. For the thorite endmember, our results ($a$ =

7.1839 Å and $c$ = 6.3511 Å) agrees excellently with previous experimental data[11] where the deviation is even minimized to 0.02 %. Further they are also very close to the experimental data of Estevenon et al., Taylor and Ewing and Guo et al.[3,12,47]. For the reidite phase of Coffinite, our results are in excellent agreement (~ 0.58%) with previous experimental[33] result. Our calculations predicted the lattice parameters of the reidite type high-pressure polymorph of Thorite: a = 5.0251 Å, c = 11.6072 Å for the first time.

Taken into account the inherent and ubiquitous overestimating trait of GGA formalism, our results are reasonably accurate and hence our predicated results for the reidite phases can further be used without caution. To the best of our knowledge there has been no previous theoretical or experimental attempt targeted towards the complete structural characterization of the reidite phases of U/ThSiO4. Fig. 2 shows the variation of unit cell volumes and *c/a* ratio normalized for members of each phases with respect to corresponding values of coffinite and its high-pressure polymorph, respectively. The incorporation of Th by substituting U enforces a linear increase in volume for both the phases, consistent with the Vegard's law. These increments of volumes are expected since the atomic radius of Th (0.24 nm) is a little larger than that of U (0.23 nm). But, the small difference between there atomic radii is chemically desirable aiding to the tolerance of the crystal structure itself for mixing ions of different sizes in a given site for the genesis of the silicate solid solution. On another note, substitution between U and Th preserves the composite electrical neutrality of the crystalline aggregate and commensurate with the local charge balance through their coordination environment in their crystallographic site. However, the manner of facilitating the volume expansion is markedly distinct for different phases appearing as obvious from the variation of *c/a* ratio (Fig. 2). The reidite phase undergoes a monotonic gain in *c/a* whereas the zircon phase exhibits a completely antipodal behavior. This result provide reinforcement to the finding of Dutta et. al.[26] who showed that the reidite demonstrates a greater compressibility along *c*-axis compared to *a*-axis and vice versa in zircon phases. Our analysis confirms that the increment along *a(c)* is more administrative in volume change in zircon (reidite) phase. An explanation from a different theoretical angle for this is provided in Section-III-C.

Since the EOS description and the comparison between zircon structured thorite and its high pressure reidite structured polymorph is scanty in literature, in Fig-3 we provided the total energy as a function of volume for both the phases. The third order Birch-Murnaghan [48] equation of state fit for both the phases are also shown. According to our literature survey, there are no previous computational or experimental attempt made in such a direction. In order to appreciate the coherence of the plot, the energy and volume of both the phases were normalized with respect to the total number of atoms present in the conventional cell. The plot makes it evident that in ambient condition the former phase is stable compared to the later.

## B. Anomalous Phase Transitions and Bulk Moduli

Deformation induced strain energies stored in materials play a very pivotal role in governing its behavior and is the most prolific thermodynamic parameter enforcing structural phase transition. We theoretically estimated the effect of hydrostatic pressure on the variations of volumes of compositionally different members in both zircon and reidite phases. The probability of a transition is determined by the height of the energy barrier and by the ability of the system to overcome it. To obtaining the critical hydrostatic pressure required for zircon type to reidite type structural transition between we have used the formalism of enthalpy crossover of the compositionally identical members of the solid solutions. We have evaluated the ground state enthalpies of the structures to determine the required hydrostatic pressure for transition. However, the enthalpy equality is a necessary and sufficient condition for structural phase transition at T = 0K. At the transition pressure $p=p_t$ both structures are mechanically stable.

Our study reveals a peculiar relation of $p_t$ with the concentrations of the different heavy radionuclide e.g. U and Th (Figure-4). As it is evident from values of $p_t$ presented in Table I and Fig. 4, $p_t$ starts from a maximum value of 8.52 GPa for the U-only end-member which begins to decline as 25 at% of U is replaced by the Th ($p_t$ = 7.67 GPa). The transition pressure attains a minimum of 6.82 GPa at Th and U occur in equal concentration. Further substitution reverses the track of transition pressure variation and it increases again finally attaining a value of 8.68 GPa for $ThSiO_4$, a value comparable to $USiO_4$. Our calculated transition pressure for $USiO_4$ demonstrates deviations from earlier experiments[32,33]. The pressure transmitting mediums chosen in experiments by Bauer et al. and Zhang et al. are neon and (16/3/1) methanol/ethanol/water mixture respectively. The pressure transmitting medium is selected in such a way that the medium should behave hydrostatically within the operating range of pressure. But the methanol/ethanol/water mixture and neon exhibits signature of nonhydrostacicity at 10-11 GPa and 15 GPa[49] respectively. Beyond that range of pressures the media start to develop and support shear while deviating from hydrostacicity, which may introduce errors in accurately determining the transition pressure. Besides, in ab-initio calculations the value of transition pressures depends on the choice of pseudopotenial and the treatment of exchange-correlation effect in them to a considerable extent and last but not the least, on temperature. These are the main reasons for the deviation of $p_t$ that we have determined from the earlier reported values. However, our calculated $p_t$ for $ThSiO_4$ is consistent with force field study of Bose et al.[25] who predicted a transition pressure greater than 3 GPa. Fitting the E-V curve of coffinite with third order Birch-Murnaghan equation of state yields a bulk modulus of 181.3 GPa, which is in excellent agreement with previous values of 181 GPa[33] and 188 GPa[32]. To the best of our knowledge the bulk moduli of the Th-end member and other interim members of zircon type phases are not reported elsewhere. Table I makes it evident that while the reidite type phases show a monotonic decrease in bulk moduli with increasing Th content and volume of unit cell, the bulk moduli of zircon type phases behave in a distinct manner. The end-members of the zircon type series have comparable values of bulk moduli but any departure from pure phases reduces the values. The lowest value is obtained when Th at% = U at%. The departure of the bulk moduli in case of high pressure $USiO_4$ may be attributed to the previously mentioned nonhydrostacicity of the pressure transmitting medium.

The volume collapses in these transitions are constrained within 10.38% to 10.57%. It is least for coffinite and maximum for thorite. Bauer et al. have reported an experimental volume collapse of 11.18% for coffinite which is confirmed to a very good extent by our theoretical study. In Figure-5 we have shown the variation of compressibilities for both phases with respect to concentration of Th. It is clear that the compositional midmember of the zircon type phase has the highest compressibility implying that $U_{0.50}Th_{0.50}SiO_4$ is the most pressure sensitive among them. Consequently, a lower pressure is capable enough to trigger the solid state structural transition. The values of compressibility of reidite type phases increase monotonically as expected from the bulk moduli data from Table I.

## C. Measurement of Distortions of U/Th polyhedron

### i. Definition

Actinide centered eight-fold coordination polyhedron having the shape of snub-disphenoid can be found in different structure types, such as, Monoclinic- $CuTh_2(PO4)_2$[50], Trigonal- Dugganite[51], Orthorhombic- Vitusite and $CaU(PO4)_2$[52],Tetragonal- Th-bearing Orthoarsenates,

hexagonal- rare earth and Th-bearing hydroxyl phosphate or Rabdophane group of minerals. Hence, a competent analysis of the geometrical peculiarity of snub-disphenoids merits a discussion and is capable of providing new structural insights. In our study, we have encountered the perplexing geometry of polyhedra which are at the same time irregular and distorted in shape. In this section, we defined two physical parameters to quantify the distortion and have shown how these distortions can have a promising role to oversee the structural transition.

For ideal regular polyhedra, there is always a unique point inside it, which is equidistant from the vertices. It denotes the position of the cation center and the vertices serve as the positions of the anions or the central atom of the ligands in coordination complexes. The length between the unique point and the vertices is analogous to ideal and unique bond length. This regularity in arrangement requires the angle formed at the center by any two adjacent vertices to be unique in undistorted polyhedra. The distortions are then quantified in comparison to the ideal unique bond length and the bond angles formed at the center[35,36]. But it has been observed that even in ideal but irregular polyhedra the center to vertex length i.e. the bond length and the bond angle morphs into a number of lengths and angles with different multiplicities. Consequently, the existing methods deems impractical for quantification of longitudinal or angular distortion.

A critical inspection of the conformational skeleton of a snub-disphenoid presents us with an avant-garde approach. The U/Th atom is 8-coordinated with O atoms which manifests themselves as an irregular geometric structure called triangular dodecahedra or mathematically popular as snub-disphenoid. An ideal snub-disphenoid has 8 vertices (i.e. the locations of the O atoms in our case) and 12 equilateral triangles as its enclosing surfaces. The line between two adjacent O-atoms forms 18 edges of the snub-disphenoid. However, all the vertices are not equivalent. Four of a kind form the edge by assembling with four nearest neighbors and the rest four of the other kind of vertices do so with five nearest neighbors (see Fig. 1 and Fig. 6). We designate the four O atom positioned at the former locations as $O_4$-s and at the later as $O_5$-s. To add to this, there is another peculiarity regarding the heterogeneity in the type of oxygen atoms in those U/Th polyhedron. In case of zircon phases, wherever the U/Th polyhedra share an edge with the Si-tetrahedra, the resulting edge contains two $O_4$-s in its extremities contrary to point sharing with Si-tetrahedra, where the O-atom is of $O_5$-kind. In the reidite phases the absence of edge sharing withdraws the constraint imposed on the preference of O-atom in different geometrical sites as observed in the former case and thus allows the shared O-atom to be either $O_4$ or $O_5$.

There is a special point (X in Fig. 6) inside the chassis of the snub-disphenoid which is equidistant from the four vertices that are connected to four nearest neighbor vertices i.e. from positions of $O_4$. Let this length be $l_4$. Similarly X is also equidistant from remaining four of the vertices that are connected to five nearest neighbor vertices i.e. positions of $O_5$, with a length say $l_5$. But $l_4 \neq l_5$. Here one must keep in mind that these types of connection are purely of geometrical nature which are necessary to form the edges of snub disphenoid, not to be confused with an electronic correspondence resulting bond formation. In order to quantify the longitudinal and angular distortions we need to find $l_4, l_5$ and the different angles formed by adjacent vertices at the position of X in cases of ideal snub-disphenoids isovolumetric with our calculated ones. Let us assume an ideal snub-disphenoid of unit edge length as illustrated in Figure-6. The solutions to the coordinates are available in Wolfram Math library however we performed it independently. The answer involves solving a set of four simultaneous quadratic equations, obtained from equating the algebraic expression of each edge lengths to 1. Hence we have 8 equations in hand with 4 of them being redundant for solving, but serves as a tool to check our solutions. The solutions for p, q, r and s obtained from our calculation are given below:

p ≈ 0.64458, q ≈ 0.57837,

r ≈ 0.98949 and s ≈ 1.56786.

To find the coordinate of the central point in order to measure the ideal bond length, we adopted the Nelder-Mead optimization scheme. It involved a search for the coordinates of the center for which the sum of the lengths of the vertices from X converges to a global minimum. We obtained the coordinates of the point X as (0, 0, ~ 0.78393), followed by $l_4 \approx 0.929809$ *and* $l_5 \approx 0.676568$. A closer introspection reveals an insightful fact that the z-coordinate of the special point is *s/2* and/or *(p+r)/2*. Hence, the problem of finding the coordinate and vertices of an ideal snub-disphenoid with edge length *a*, reduces to a scaling problem. The prescription is to multiply *p, q, r, s* and the coordinate of the special point by *a*. The ideal edge length of a snub-disphenoid can be calculated from its volume via the relation

$$a \approx \left(\frac{V}{0.8594937}\right)^{1/3} \quad (1)$$

It is needless to say that different set of $l_4$ and $l_5$ can also be found for ideal snub-disphenoids with different volume following the same way. The striking observation in this case is that the ratio of $l_4$ to $l_5$ remains a constant *L* (≈ 1.374302), irrespective of the volume of the ideal snub-disphenoid. A similar ratio can be found for distorted U or Th snub-disphenoids as $l_U$ *and* $l_{Th}$ respectively. So, a deviation of $l_U$ *and* $l_{Th}$ from *L* can be a measure of distortion of the polyhedra under study. We define

$$\delta_U = (l_U - L)^2 \quad and$$

$$\delta_{Th} = (l_{Th} - L)^2 \quad (2)$$

as a measure of longitudinal distortions of U and Th-polyhedra respectively.

Switching our argument to angles, one can think of bond angles formed by any two adjacent vertices at X so that the edges of the polyhedra will be opposite to the bond angle. But since the vertices are of two kinds, the adjacent vertices may or may not be of the same kind. This contrast can give rise to three geometrically and numerically distinct angles of multiplicity 2, 4 and 12 respectively: $O_4 - X - O_4(\alpha), O_5 - X - O_5(\beta)\ and\ O_4 - X - O_5(\gamma)$. In ideal cases we find that $\alpha = 65.061°\ , \beta = 95.297°\ and\ \ \gamma$ = 75.158° from the laws of cosines. We can then consider three different sample angular variances $\sigma_\alpha^2, \sigma_\beta^2\ and\ \sigma_\gamma^2$ or rather their sum $\sigma^2$. Thus, according to our definition

$$\sigma_\alpha^2 = \sum_{i=1}^{2}(\alpha_i - 65.061)^2$$

$$\sigma_\beta^2 = \frac{1}{3}\sum_{j=1}^{4}\left(\beta_j - 95.297\right)^2]$$

$$\sigma_\gamma^2 = \frac{1}{11}\sum_{k=1}^{12}(\gamma_k - 75.158)^2$$

And, finally $$\sigma^2 = \sigma_\alpha^2 + \sigma_\beta^2 + \sigma_\gamma^2 \qquad \text{……………(3)}$$

In the following subsection we describe the unprecedented influence these distortions have on the anomalous phase transition pressures.

### ii. Analysis: Connection to Phase Transitions

The previous discussion explains with reason how co-ordination geometry is responsible for the evolution of two types of U/Th-O bond lengths each with multiplicity 4. Table 2 summarizes our calculated bond lengths along with pre-existing data. Our calculated U-$O_4$ and U-$O_5$ bond-lengths for coffinite (2.438 and 2.311 Å) are within 0.8% of both the synchroton powder measurement and DFT+U study of Bauer et.al[33]. Our calculated values are in excellent agreement with the values from EXAFS measurement by Labs et. al.[11] (2.439 and 2.298 Å i.e. within 0.5 %). For the zircon type $ThSiO_4$ our calculated Th-$O_5$ bond length (2.469 Å) is very much consistent with the results of Fuchs and Gebert[16] (2.47 Å) and Labs et al.[11] (2.467 Å). The shorter bond length (2.384 Å) shows a little deviation from the previous results of Fuchs and Gebert and Labs et al. As it reveals, $O_5$-s are closer to the central U/Th atom than the $O_4$-s. In zircon phase the length of U/Th-$O_4$ bonds are negligibly responsive to stoichiometric exchange between U and Th. But a monotonic increment is observed in the length of U/Th-$O_5$ bonds which are especially instrumental in volume expansion due to increasing Th-content. In case of the high pressure reidite type $USiO_4$ the longer U-$O_4$ bond length is appreciably closer to the DFT+U result of Bauer et al.[33] and force-field calculation of Bose et.al[25]. The shorter U-$O_5$ bond length (2.344 Å) is within 1.8 % of both the previously mentioned reports. It is evident that stoichiometric interchange between U and Th effects both the bond lengths in similar fashion and induces very little change in both them. The relatively open unit cell of reidite phases permits them to accommodate Th in place of U by virtue of a small volume change without altering the bond lengths significantly. The $SiO_4$ tetrahedras, being more rigid than the U-Th polyhedras does not undergo significant geometrical modifications and the length of Si-O bonds remains constrained well within 1.635 Å to 1.667 Å which are commensurate with the references. In Fig. 2 we have observed a monotonic decrease in *c/a* ratio corresponding to increasing Th-content and unit cell volume in zircon type phases. Here we like to elaborate that the U/Th-$O_5$ bonds lie almost parallel to *a* and *b*-axes. So, the increment in their lengths facilitates the comparatively larger increment in lattice parameters along *a* and *b*-axes. Whereas, the U-$O_4$ bonds does not have any direct influence on the length of lattice parameter along *c*. This is the reason for the pattern of *c/a* alteration in zircon type phases. The length of the U/Th-$O_4$ does not exhibit significant changes because they form the edge which is shared by both the U/Th and Si-polyhedra giving rise to a very strong connection between the U/Th or Si atom and O atom. In contrary, U/Th-$O_5$ bond undergo notable changes because $O_5$'s are being shared between U/Th polyhedra and Si-tetrahedra as an atom only, causing the connections to be weaker compared to the U/Th-$O_4$ bonds. (See Fig-1). In reidite type phases no preferred orientation of bonds with respect to the axis of the unit cell is observed. Both the U/Th-$O_4$/$O_5$ bonds take part in volume increment followed by *c/a* increment with amount of Th content.

Table-III sums up the calculated values of $\delta_U$ , $\delta_{Th}$ , $\sigma_U^2$ $and$ $\sigma_{Th}^2$ according to our definitions in Eq. 2 and 3. Our analysis confirms that elemental interchange between U and Th has profound impact on the length of the corresponding bonds and the distortion of the polyhedron formed by these bond dictates the specific hydrostatic pressure required for the zircon to reidite type transitions. Their variations with concentration of Th in the unit cell is demonstrated in Fig. 7. The distortions of bond lengths in both Uranium ($\delta_U$) and Thorium ($\delta_{Th}$) polyhedra in zircon type phase increases with increase in Th %. But the corresponding variations in reidite type phases exhibits an increment up to the point where the concentration of U and Th are equal (both are 50 %). Further gain in Th percentage results in a decline of the distortion of the bond. These trait asserts that the change in bond length distortion of both U and Th-polyhedron in reidite type phases is characterized by the presence of two inherent thresholds on the maximum side. The numerical values of these thresholds are $\delta_U = 0.12223$ and $\delta_{Th} = 0.11478$. The angular distortion of U-O bonds, $\sigma_U^2$ on the other hand decreases (increases) in reidite (zircon) type phases until the amount of Th content becomes equal with the U content where the slope of the curves changes its sign (Fig. 7a). A similar, although quantitatively different, trend is observed in case of angular distortion of Th-O bonds represented by $\sigma_{Th}^2$(Fig. 7b). The variation $\sigma_U^2$ $and$ $\sigma_{Th}^2$ are also characterized by an indicative maximum (minimum) limit in zircon (reidite) type phases. Evidently, the numerical

values of these limits of bond length and angular distortions that these phases allow are dictated solely by the competing amount of the constituent Actinides. As the hydrostatic pressure is increased these zircon type phases try to reorient and reorganize the polyhedral structure in order to overcome the corresponding distortions. According to the Fig. 7, it is eminent that the difference between bond angular distortions between zircon and reidite type phases is minimum when U and Th percentage are equal. Correspondingly the pressure required to trigger the transition attains a minimum (6.82 GPa). In fact, the increment or diminution of Th percentage from 50%, stimulates an increment $\sigma_U^2\ and\ \sigma_{Th}^2$ in reidite type phases and a reduction in zircon type phases. These alterations raise the difference of the distortions between stoichiometrically similar zircon and reidite type phases as an effect of which we observe a larger pressure of transition. Since these reconstructive phase transitions proceeds through an intermediate transient cubic phase[31], our theoretical investigation predicts that the difference between the numerical estimates of the distortions in the corresponding zircon and reidite type phases is a factor of paramount importance which in turn influences and fixes required critical hydrostatic pressure of transition.

In Fig. 8 we have provided a quantitative description of how $\delta_U$ and $\delta_{Th}$ varies with applied hydrostatic stresses in the zircon phase. As it can be seen that both $\delta_U$ and $\delta_{Th}$ decrease with increasing pressure. However, it is quite interesting to note that $\delta_U$ for USiO4 is the lowest in ambient condition and it increases as more and more Th is substituted in place of U suggesting that even the deformation behavior of U-polyhedra is in part controlled by the larger element Th. In case of Th-polyhedra, $\delta_{Th}$ is maximum for pure $ThSiO_4$ and it decreases with lowering of Th content. So, in summary elemental Th has larger influence on the polyhedral distortion than elemental U. Our analysis shows that the slope of both $\delta_U$ and $\delta_{Th}$ are found to decrease as the pressure increases. The lesser value of $\delta_U$ and $\delta_{Th}$ at high pressures are suggestive of the fact that the pressure aids to lower the inherent distortion behavior and pushes it towards an ideal polyhedral structure. At the high pressure regime (beyond 10 GPa) the average of the slope for $\delta_U$ i.e. $\frac{\Delta\,\delta_U}{\Delta P}$ in $USiO_4$ and $\delta_{Th}$ in $ThSiO_4$ i.e. $\frac{\Delta\,\delta_{Th}}{\Delta P}$ are found to be 0.001265 /GPa and 0.001324 /GPa, respectively.

We also tried to plot the same for the reidite phases. They exhibited similar kind behavior in response to pressure however the slopes of their variation are much lower compared to what is observed in zircon phases. Hence, in reidite phases the effect of pressure on the distortion is not as eminent as in zircon phases. In order to $\delta_U$ and $\delta_{Th}$ to become zero, which is the case of ideal snub-disphenoid, a much higher pressure is needed in reidite phases. On another note, the dependencies of $\delta_U$ and $\delta_{Th}$ on amount of elemental Th is absent in reidite phases. The absence of edge sharing between the U/Th-polyhedra and Si-tetrahedra and the presence of large amount of empty spaces in the conventional unit cell of reidite phases remove the constraint imposed on them from structural point of view. As a result of which, their distortion behavior w.r.t. pressure is not as per with the zircon type phases.

## List of Figures with captions:

## Figure-1:

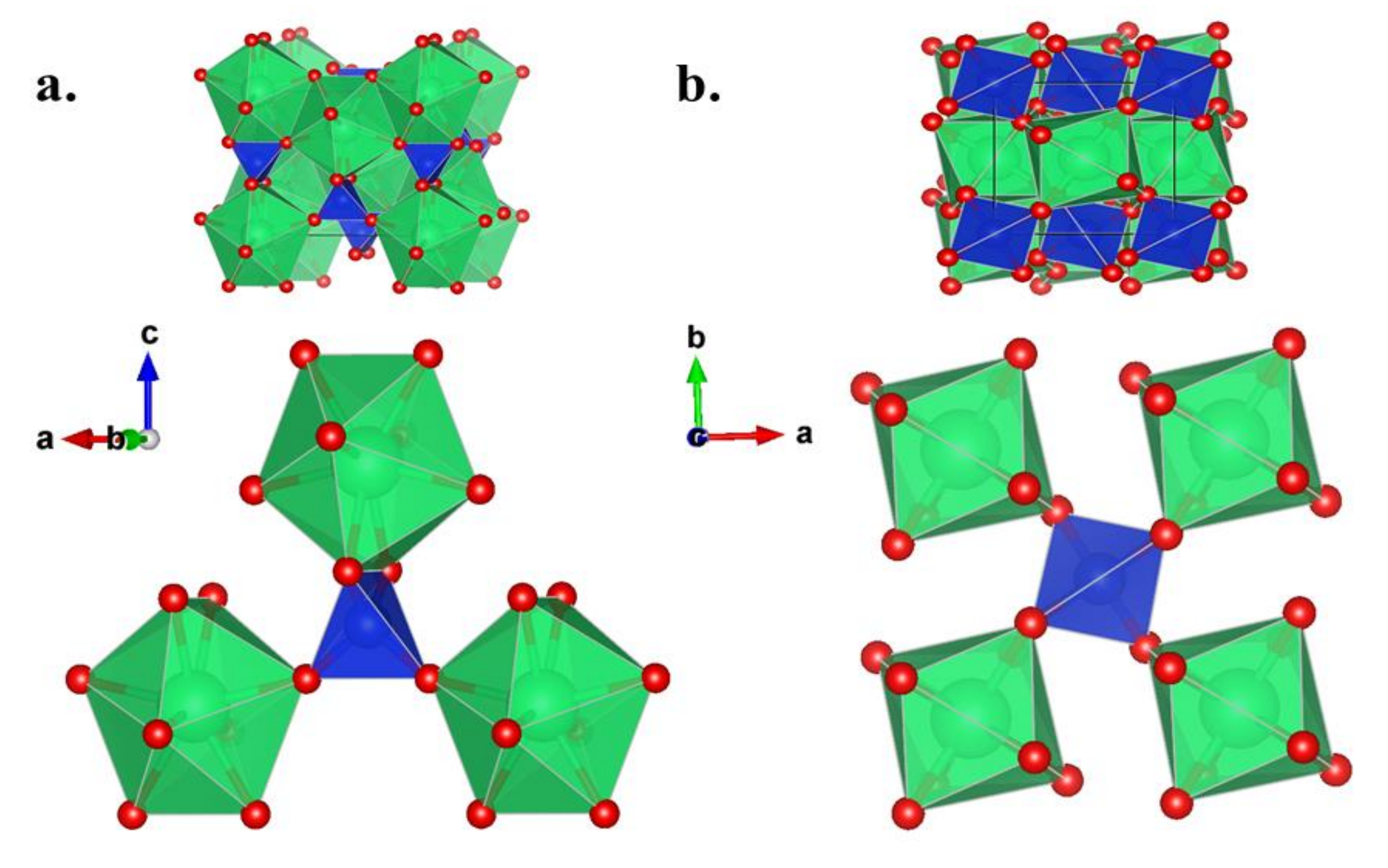

**Figure-1**: The unit cells of $ThSiO_4$ and the connection and alignment of the Th-dodecahedra with respect to Si-tetrahedra in **a.** zircon type phase and **b.** reidite type phase. Green polyhedras are Th-polyhedra and blue polyhedras are Si-tertrahedras. Red spheres are Oxygen atoms. The reidite structure is comparatively open with a hollow tunnel parallel to crystallographic c-axis.

**Figure-2:**

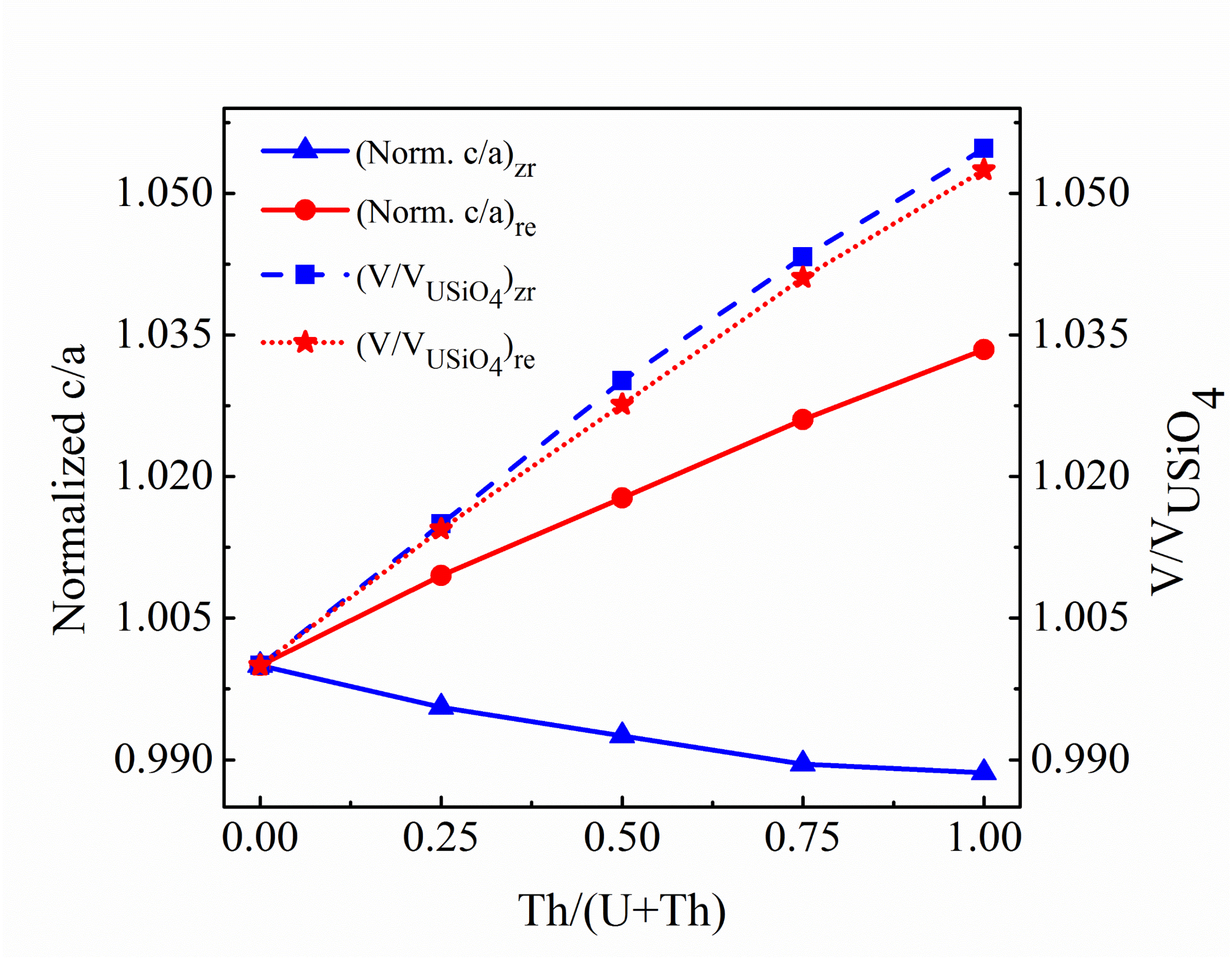

**Figure-2:** Variation of c/a ratio and unit cell volume of $U_{1-x}Th_xSiO_4$ in both zircon and reidite type phases, where x = Th/(U+Th). For a better comparison all the c/a ratios and volumes have been normalized with respect to the corresponding values of $USiO_4$ member of the corresponding phase. Notice the distinct nature of undergoing volume change in different phases: c/a increment in reidite and opposite in zircon.

**Figure-3:**

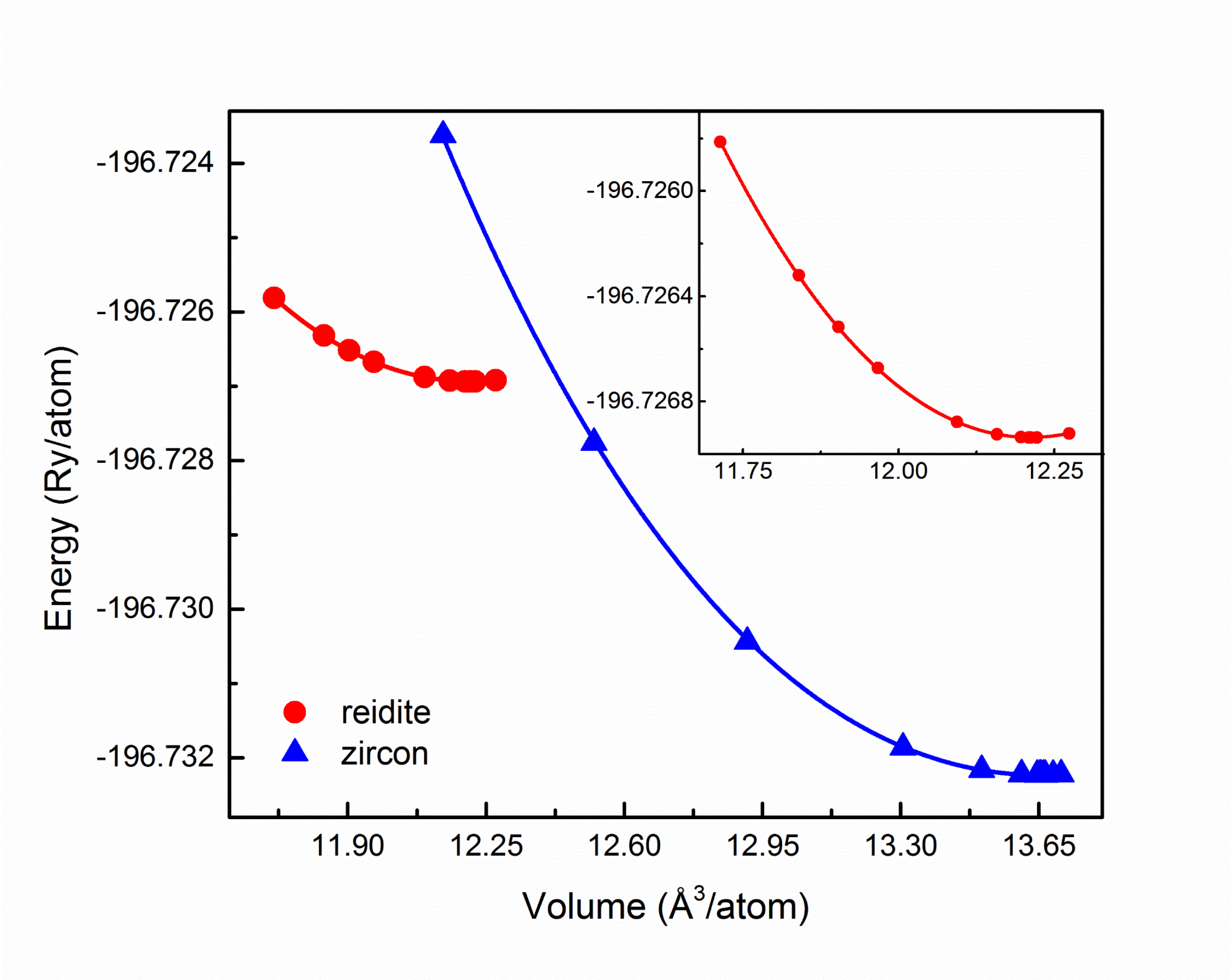

**Figure-3:** 3rd order Birch-Murnaghan's equation of state fit for zircon-type $ThSiO_4$ and its' reidite polymorph. The inset shows an enhanced view of the EOS fit of reidite-type $ThSiO_4$. We confirm that no data point has been obscured by the inset graph.

**Figure-4:**

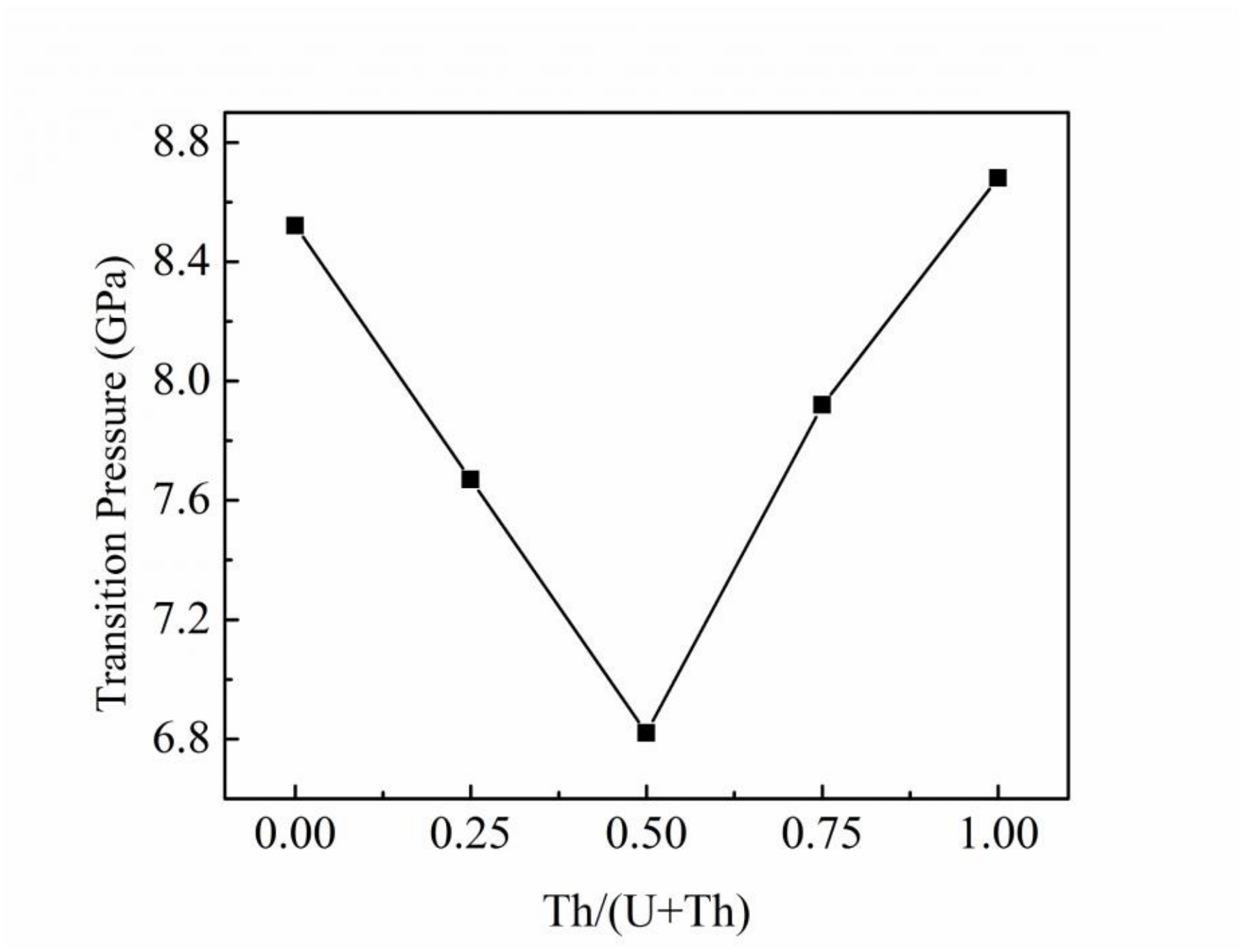

**Figure-4:** Variation of Zircon→reidite type transition pressure of $U_{1-x}Th_xSiO_4$, x = Th/(U+Th). The transition pressure for end members of this stoichiometric spectrum are almost equal (~ 8.6 GPa). The transition pressure attains a minimum when the concentrations of U and Th are equal.

**Figure-5**

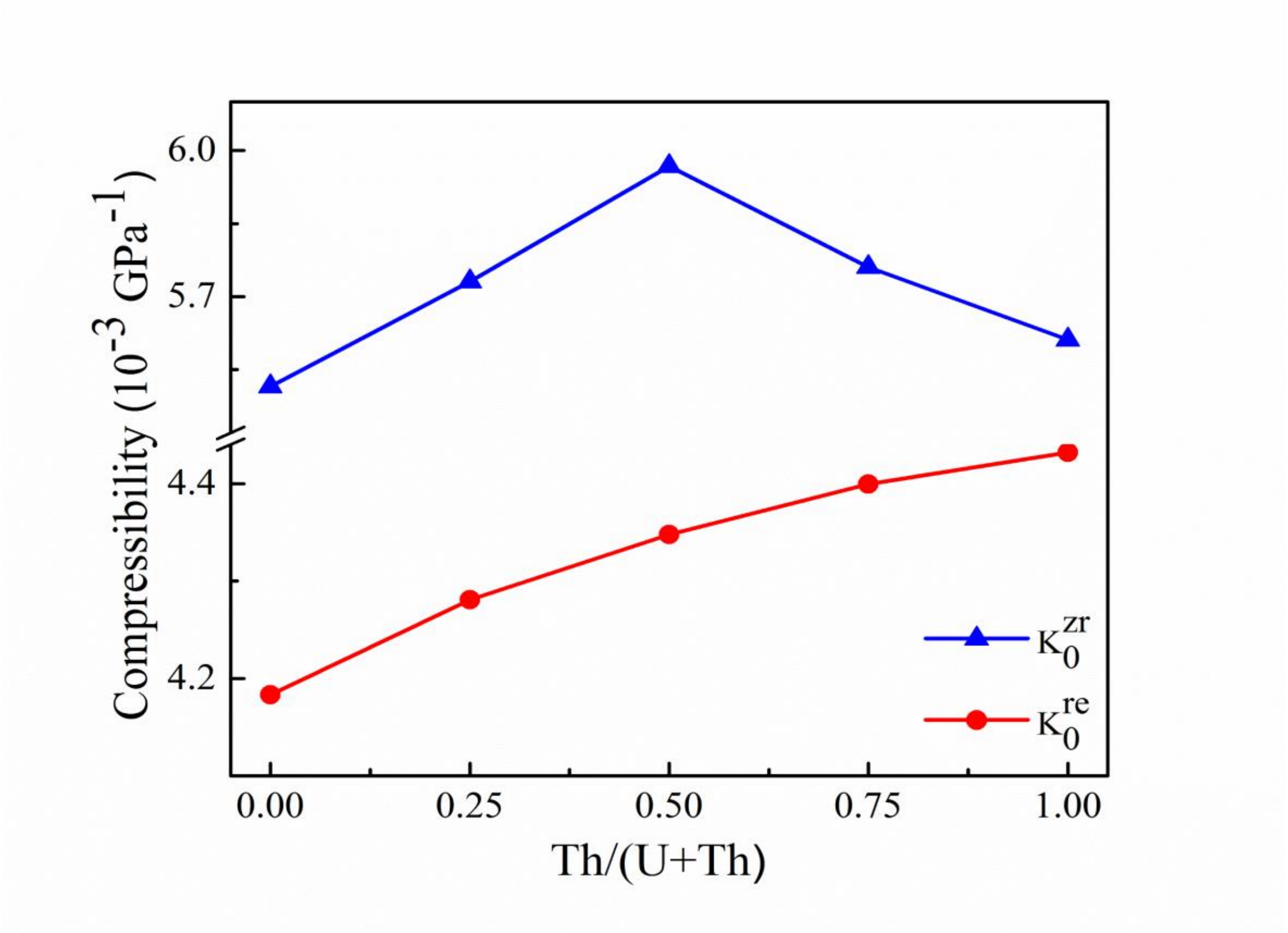

**Figure-5:** Compressibilities of $U_{1-x}Th_xSiO_4$, x = Th/(U+Th) in both phases.

**Figure-6**

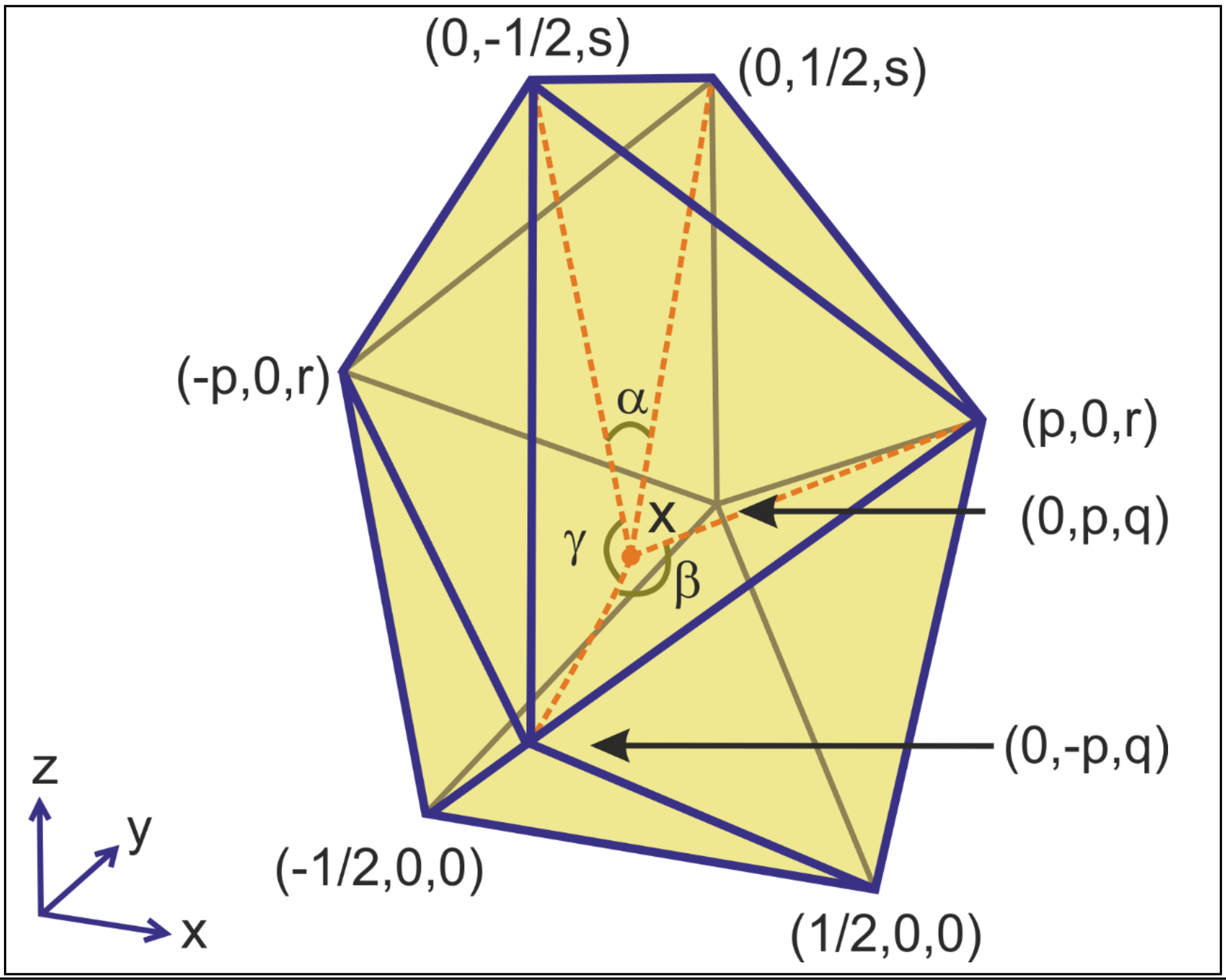

**Figure-6:** Distribution of coordinates of vertices of an ideal snub-disphenoid of unit edge length. The origin lies at the midpoint of the bottom edge, as evident from the chosen coordinates of the vertices. The red-point (X) inside is the special point as discussed above. The vertices with p and q in their coordinates are the position of $O_5$ and the rest are the position of $O_4$. The different type of bond angles that can occur with different multiplicities are also shown as α, β and γ.

**Figure-7:**

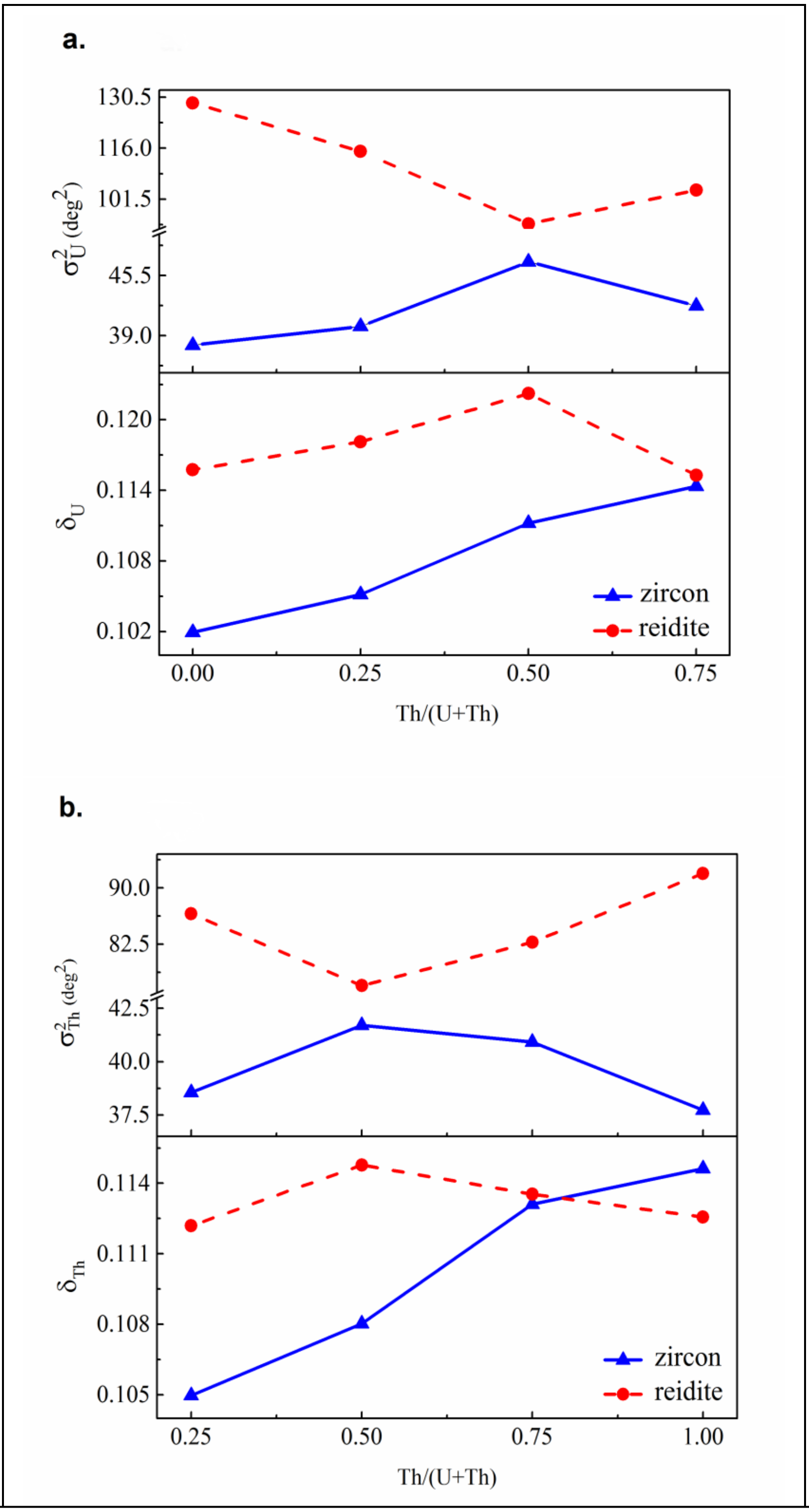

**Figure-7:** Variation of distortions of **a.** Uranium-snub disphenoid, **b.** Thorium-snub disphenoid in zircon and reidite-type $U_{1-x}Th_xSiO_4$ with respect to x = Th/(U+Th) ratio. Distortions are quantified in terms of bond angle variance ($\sigma^2$) and bond length ratio contrast ($\delta$).

**Figure-8**

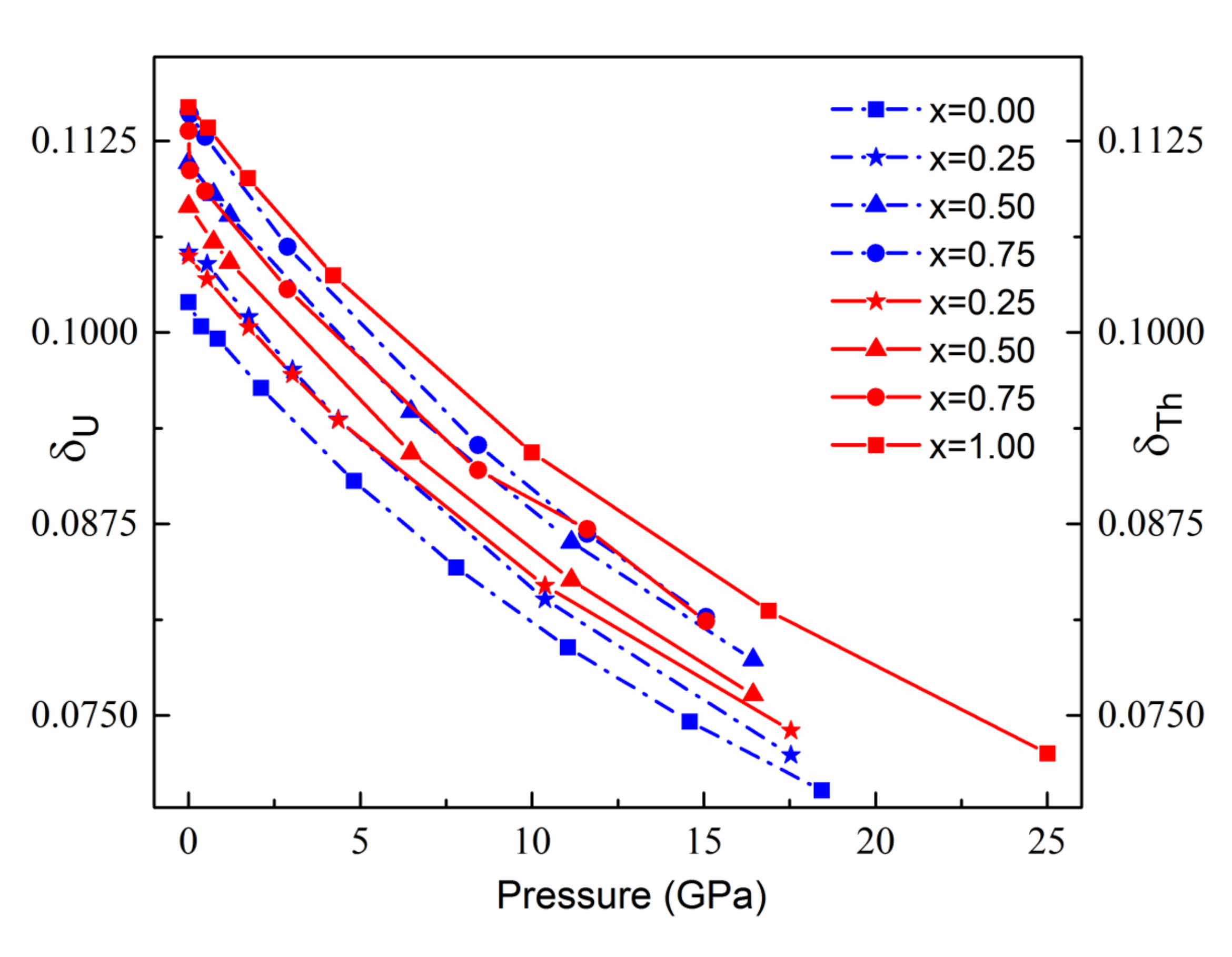

**Figure-8:** The pressure dependent variation of $\delta$ for both U and Th-polyhedra. The blue lines and symbols are for Uranium and the red lines and symbols are for Thorium. The x in legend stands for x = Th/(U+Th) like the previous cases. Hence, the square symbols characterize the pure U end-member $USiO_4$ (red) and pure Th end-member $ThSiO_4$ of $U_{1-x}Th_xSiO_4$ series.

# List of tables with captions:

**Table-I**

| Phase (Sp. Gr.) | x | a (Å) | c (Å) | c/a | Volume (Å$^3$) | Bulk Modulus (GPa) | Transition Pressure (GPa) |
|---|---|---|---|---|---|---|---|
| Zircon (I4$_1$/amd) | 0.00 | 7.0303 | 6.2872 | 0.8942 | 310.75 | 181.30 | - |
| | | 7.0135* | 6.2669* | | - | - | |
| | | 6.981† | 6.250† | | - | - | |
| | | 6.9862‡ | 6.2610‡ | | 305.58‡ | 181‡ | |
| | | 6.9842§ | 6.2606§ | | 305.38§ | - | |
| | | 6.9904‖ | 6.2610‖ | | 305.94‖ | - | |
| | | 6.9936¶ | 6.2614¶ | | 306.25¶ | 188¶ | |
| | 0.25 | 7.0758 | 6.2997 | 0.8904 | 315.41 | 174.49 | - |
| | | 7.0105§ | 6.2680§ | | 308.06§ | - | - |
| | | 7.007† | 6.275† | | | | |
| | 0.50 | 7.1181 | 6.3180 | 0.8876 | 320.11 | 167.57 | - |
| | | 7.039† | 6.294† | | | | |
| | 0.75 | 7.1553 | 6.3320 | 0.8849 | 324.19 | 173.60 | - |
| | | 7.0949§ | 6.3194§ | | 318.10§ | - | - |
| | | 7.071† | 6.314† | | | | |
| | 1.00 | 7.1839 | 6.3511 | 0.8841 | 327.75 | 178.20 | - |
| | | 7.1816§ | 6.2946§ | | 324.66§ | - | |
| | | 7.1568‖ | 6.3152‖ | | 323.46‖ | - | |
| | | 7.133** | 6.319** | | - | - | |
| | | 7.1328†† | 6.3188†† | | 321.48†† | - | |
| | | 7.1439‡‡ | 6.3183‡‡ | | 322.46‡‡ | - | |
| | | 7.129§§ | 6.319§§ | | 321.15§§ | | |
| | | 7.128† | 6.314† | | | | |
| Reidite (I4$_1$/a) | 0.00 | 4.9794 | 11.2318 | 2.2557 | 278.49 | 239.05 | 8.52 |
| | | 4.9502‡ | 11.0750‡ | | 271.39‡ | 212‡ | 15‡ |
| | | 4.8454¶ | 11.0316¶ | | 261.14¶ | 274¶ | 14-17¶ |
| | 0.25 | 4.9915 | 11.3389 | 2.2717 | 282.51 | 233.60 | 7.67 |
| | 0.50 | 5.0053 | 11.4308 | 2.2837 | 286.18 | 230.00 | 6.82 |
| | 0.75 | 5.0157 | 11.5241 | 2.2976 | 289.92 | 227.30 | 7.92 |
| | 1.00 | 5.0251 | 11.6072 | 2.3099 | 293.10 | 225.64 | 8.68 |

* V. Pointeau et al., J. Nucl. Mater., 393, 3 (2009)

† L.H. Fuchs and E. Gebert, Am. Min., 43, 3-4 (1958)

‡ J. D. Bauer et al., J. Phy. Chem. C, 118, 43 (2014)

§ S. Labs et al., Environ. Sci. Technol., 48, 1 (2014)

‖ X. Guo et al., Chem. Mater., 28, 19 (2016)

¶ F. X. Zhang et al., Am. Min., 94, 7 (2009)

** I. R. Shein et al., Phy. Chem. Min., 33, 8-9 (2006)

†† M. Taylor and R. C. Ewing, Acta Cryst., B34, 1074-1079(1978)

‡‡ P. Estevenon et al., Inorg. Chem., 57, 15 (2018)

§§ L. Mazeina et al., Geo. Cosmo. Acta., 69, 19 (2005)

**Table-I.** Optimized lattice parameters, c/a ratios, volumes of unit cells, bulk moduli for $U_{1-x}Th_xSiO_4$ solid solutions in both ambient pressure zircon phase and its high pressure polymorphic phase viz reidite. The last column enlists the hydrostatic pressures required for zircon→reidite phase transition between stoichiometrically identical members of the series.

## Table-II

| *Phase* | *x* | *U-0$_4$ (Å)* | *U-O$_5$ (Å)* | *Th-0$_4$ (Å)* | *Th-O$_5$ (Å)* | *Si-O(m) (Å)* |
|---|---|---|---|---|---|---|
| | 0.00 | 2.438 | 2.311 | - | - | 1.647(4) |
| | | 2.418* | 2.310* | | | |
| | | 2.439† | 2.298† | | | |
| **Zircon** | 0.25 | 2.440 | 2.324 | 2.465 | 2.347 | 1.660(2),1.635(2) |
| | 0.50 | 2.438 | 2.342 | 2.469 | 2.362 | 1.651(2),1.647(2) |
| | 0.75 | 2.442 | 2.357 | 2.464 | 2.374 | 1.655(2),1.647(2) |
| | 1.00 | - | - | 2.469 | 2.384 | 1.650(4) |
| | | | | 2.467† | 2.363† | |
| | | | | 2.47‡ | 2.37‡ | |
| | 0.00 | 2.424 | 2.344 | - | - | 1.660(4) |
| | | 2.4304* | 2.3895* | | | |
| **Reidite** | | 2.474§ | 2.328§ | | | |
| | 0.25 | 2.421 | 2.349 | 2.479 | 2.385 | 1.665(2),1.656(2) |
| | 0.50 | 2.412 | 2.354 | 2.476 | 2.391 | 1.667(2),1.658(2) |
| | 0.75 | 2.441 | 2.359 | 2.483 | 2.393 | 1.662(2),1.657(2) |
| | 1.00 | - | - | 2.487 | 2.394 | 1.659(4) |

* J. D. Bauer et al., J. Phy. Chem. C, 118, 43 (2014)

† S. Labs et al., Environ. Sci. Technol., 48, 1 (2014)

‡ Taylor, M.; Ewing, R. C., Acta Crystallogr. 34 (1978)

§ Bose, P. P.; Mittal, R.; Chaplot, S. L., Phys. Rev. B: Condens. Matter Mater. Phys. 79 (2009)

**Table-II.** Calculated bond lengths of $U_{1-x}Th_xSiO_4$ in both zircon and reidite phases with varying U and Th percentage. The bracket in the last columns succeeding the data are the multiplicity of that bond.

**Table-III:**

| Phase | x | $\delta_U$ | $\delta_{Th}$ | $\sigma_U^2$ ($deg^2$) | $\sigma_{Th}^2$ ($deg^2$) |
|---|---|---|---|---|---|
| **Zircon** | 0 | 0.10198 | - | 37.982 | - |
| | 0.25 | 0.10523 | 0.10497 | 40.007 | 38.555 |
| | 0.50 | 0.11110 | 0.10824 | 46.984 | 41.707 |
| | 0.75 | 0.11441 | 0.11316 | 42.226 | 40.923 |
| | 1.00 | - | 0.11468 | - | 37.732 |
| **Reidite** | 0.00 | 0.11572 | - | 128.891 | - |
| | 0.25 | 0.11810 | 0.11215 | 115.069 | 86.555 |
| | 0.50 | 0.12226 | 0.11475 | 94.471 | 76.962 |
| | 0.75 | 0.11523 | 0.11336 | 104.038 | 82.762 |
| | 1.00 | - | 0.11253 | - | 91.937 |

**Table-III:** Distortions of bond lengths and bond angles of U and Th snub-disphenoids in compositionally different zircon type structures of $U_{1-x}Th_xSiO_4$ and in their corresponding reidite phase. The subscript in the name of the parameters in heading refers to actinide atomic species at the point X inside the snub-disphenoid as shown in Fig. 6.